\documentclass{article}%
\usepackage{amssymb}
\usepackage{amsfonts}
\usepackage{amsmath}%
\usepackage{graphicx}
\providecommand{\U}[1]{\protect\rule{.1in}{.1in}}
\begin{document}

\title{Electrodynamic Radiation Reaction and General Relativity}
\author{Carlos Kozameh$^{1}$, Ezra T. Newman$^{2}$, Raul Ortega$^{3}$ Gilberto
Silva-Ortigoza$^{4}$\\$^{1}$FaMaF, Univ. of Cordoba, \\Cordoba, Argentina\\kozameh@famaf.unc.edu.ar\\$^{2}$Dept of Physics and Astronomy, \\Univ. of Pittsburgh, \\Pittsburgh, PA 15260, USA\\newman@pitt.edu\\$^{3}$FACEN, Univ. of Catamrca, \\Catamarca, Argentina\\nikkoortega@yahoo.com.ar\\$^{4}$Facultad de Ciencias F\'{\i}sico Matem\'{a}ticas \\de la Universidad Aut\'{o}noma de Puebla, \\Apartado Postal 1152, 72001,\\Puebla, Pue., M\'{e}xico\\gsilva@fcfm.buap.mx}
\maketitle

\begin{abstract}
We argue that the well-known problem of the instabilities associated with the
self-forces \ (radiation reaction forces) in \ classical electrodynamics are
possibly stabilized by the introduction of \ 

gravitational forces via general relativity.

\end{abstract}

\section{Introduction}

The problems and difficulties associated with the motion of charged particles
interacting with both an external electromagnetic field and its own self-field
have not been resolved even after over a century of
investigation\cite{1,2,3,4,5}. The problems arise in many different contexts:
the difficulties in giving appropriate initial conditions, infinite
self-energy problems, model building, Lorentz invariance difficulties and
perhaps the most serious, the instabilities (or pre-acceleration) in the
solutions to the equations of motion.

The best known equations of motion, coming from a point structureless
particle, are the Abraham-Lorentz equations and the relativist generalization
the Abraham-Lorentz-Dirac equations. They are given, respectively, by\cite{3}%

\begin{equation}
m\overrightarrow{\dot{v}}=q\overrightarrow{E}+q\overrightarrow{B}%
\times\overrightarrow{v}+\frac{2q^{2}}{3c^{3}}\overrightarrow{\ddot{v}%
},\label{2}%
\end{equation}
and\cite{2}%

\begin{equation}
m\dot{v}^{a}=qF^{ab}v_{b}+\frac{2q^{2}}{3c^{3}}(\ddot{v}^{a}+\frac{1}{c^{2}%
}v^{a}\dot{v}^{b}\dot{v}_{b}),\label{1}%
\end{equation}
where $F^{ab}$ (or $\overrightarrow{E},\overrightarrow{B}$) are an external
field and are derived by a variety of means, but always with severe approximations.

It is generally acknowledged that there are fundamental difficulties with this
issue. And there seems to be a variety of different reasons, explanations and
suggested remedies. They range from: quantum theory is the resolution to the
approximations leading to these equations are wrong or even that there is no
real problem. The author, J. D. Jackson, summarizes the situation in his well
known graduate text\cite{1} as:

``The difficulties presented by this problem touch one of the most fundamental
aspects of physics, the nature of the elementary particle. Although partial
solutions, workable within limited areas, can be given, the basic problem
remains unsolved. One might hope that the transition from classical to
quantum-mechanical treatments would remove the difficulties. While there is
still hope that this may eventually occur, the present quantum-mechanical
discussions are beset with even more elaborate troubles than the classical
ones. It is one of the triumphs of comparatively recent years (\symbol{126}%
1948 - 1950) that the concepts of Lorentz covariance and gauge invariance were
exploited sufficiently cleverly to circumvent these difficulties in quantum
electrodynamics and so allow the calculation of very small radiative effects
to extremely high precision, in full agreement with experiment. From a
fundamental point of view, however, the difficulties remain.''

The purpose of this note is to describe a totally new point of view towards
this problem. This new view is completely classical, with no reliance on
quantum theory. It however does rely heavily on the Einstein-Maxwell equations
of general relativity.

The basic situation that we address is to first consider an arbitrary compact
gravitating-electromagnetic system which is taken to be the particle whose
motion we want to describe. The system is given by local mass and charge
densities and currents. There are no external fields acting on it. It is an
isolated system with arbitrary internal degrees of fredom. The program is to
solve the Einstein-Maxwell equations in the future null asymptotic region and,
from the asymptotic field (the asymptotic Weyl and Maxwell tensors), determine
a center of mass and center of charge and their laws of motion. Aside from
conditions that the total charge $Q$ be non-vanishing and the important
requirement that the centers of mass and charge should coincide, there is no
further model building.

Since our detailed calculations, which were done in the language of the
spin-coefficient formalism, are long and complicated and have appeared
elsewhere\cite{BIG}, we will just summarize the ideas and results. Basically
the calculations are done 2$^{nd}$ order in deviations from
Reissner-Nordstrom. In the spherical harmonic expansions, with frequent use of
Clebsch-Gordon products, only terms up to the $l=2$ harmonics are kept.

\section{The Complex Center of charge: Its Identification}

The basic starting idea in this work is essentially simple. It is in the
generalizations and implementations where difficulties arise.

Starting in Minkowski space in a given Lorentzian frame with spatial origin,
the electric dipole moment $\overrightarrow{D}_{E}$ is calculated from an
integral over the (localized) charge distribution. If there is a shift,
$\overrightarrow{R}$, in the origin, the dipole transforms as%

\begin{equation}
\overrightarrow{D_{E}^{*}}=\overrightarrow{D}_{E}-Q\overrightarrow
{R}.\label{D}%
\end{equation}
If $\overrightarrow{D}_{E}$ is time dependent, we obtain the center of charge
world-line by taking $\overrightarrow{D_{E}^{*}}=0,$ i.e., from
$\overrightarrow{R}=\overrightarrow{D}_{E}/Q.$ It is this idea that we want to
generalize and extend to gravitational fields.

First, however, we want to discuss other dipole issues in flat-space. Starting
on the time-like world-line at the spatial origin, we construct the family of
future directed light-cones, $\mathfrak{C}_{0},$ and investigate behavior of
the Maxwell field in the limit as null infinity is approached, i.e., Penrose's
$\mathfrak{I}^{\mathfrak{+}}$. Using the null tetrad formalism and where the
Maxwell pair $(\overrightarrow{E},\overrightarrow{B}$ or $F_{ab})$ is replaced
by the complex vector $\overrightarrow{E}+i\overrightarrow{B},$ or more
accurately by their tetrad components, ($\phi_{0},\phi_{1},\phi_{2}),$
with\cite{NP}
\begin{align}
\phi_{0}  & =F_{ab}l^{a}m^{b}\label{MaxField}\\
\phi_{1}  & =\frac{1}{2}F_{ab}(l^{a}n^{b}+m^{a}\overline{m}^{b})\nonumber\\
\phi_{2}  & =F_{ab}\overline{m}^{a}n^{b}.\nonumber
\end{align}
The asymptotic (peeling) behavior of these fields for a compact source is
given by
\begin{align}
\phi_{0}  & =\frac{\phi_{0}^{0}}{r^{3}}+O(r^{-4})\label{BondiMax}\\
\phi_{1}  & =\frac{\phi_{1}^{0}}{r^{2}}+O(r^{-3})\nonumber\\
\phi_{2B}  & =\frac{\phi_{2}^{0}}{r}+O(r^{-2}).\nonumber
\end{align}

The vector field $l^{a}$ is the tangent field to the null geodesic generators
of the null cones $\mathfrak{C}_{0}.$ At $\mathfrak{I}^{\mathfrak{+}}$, $n^{b}
$ is the tangent field to the null generators of $\mathfrak{I}^{\mathfrak{+}}$
while ($m^{a},\overline{m}^{b}$) are (the complex conjugate pair) tangent to
the two surface, $S^{2},$ the intersection of $\mathfrak{C}_{0}\ $with
$\mathfrak{I}^{\mathfrak{+}}.$

The $r$ independent quantities ($\phi_{0}^{0},\phi_{1}^{0},\phi_{2}^{0},...)$
are functions `living' on $\mathfrak{I}^{\mathfrak{+}},$ i.e., functions of
the retarded time, $u,$ (the light-cone cuts of $\mathfrak{I}^{\mathfrak{+}}$)
and ($\zeta,\overline{\zeta}$), the complex stereographic coordinates labeling
the, $S^{2},$ generators of $\mathfrak{I}^{\mathfrak{+}}.$ The components of
their spherical harmonic decomposition,
\begin{align}
\phi_{0}^{0}  & =\phi_{0i}^{0}Y_{1i}^{1}+\phi_{0ij}^{0}Y_{2ij}^{1}%
+...,\label{harmonic decomposition}\\
\phi_{1}^{0}  & =Q+\phi_{1i}^{0}Y_{1i}^{0}+\phi_{1ij}^{0}Y_{2ij}^{0}+...,\\
\phi_{2}^{0}  & =\phi_{2i}^{0}Y_{1i}^{-1}+\phi_{2ij}^{0}Y_{2ij}^{-1}+...,
\end{align}
are the asymptotically defined multipole moments and their time derivatives.

For example, the $l=0$ harmonic component of $\phi_{1}^{0}$ is proportional to
the total source charge. For us the important quantity is $\phi_{0\,i}^{0},$
the $l=1$ component of $\phi_{0}^{0}:$ $\phi_{0\,i}^{0}$ is proportional to
the (\textit{asymptotically defined}) complex dipole moment, $\vec{D}_{C}%
=\vec{D}_{E}+i\vec{D}_{M},$ where $\vec{D}_{M}$ is the magnetic dipole moment.

The problem now is: how does the $\vec{D}_{C}$ transform under an origin shift
to an arbitrary world-line? With an origin shift there will be new light-cones
and a new null vector field, $l^{\ast a},$ obtained from the old one, $l^{a},$
by a null rotation at $\mathfrak{I}^{\mathfrak{+}}.$ This can be expressed
explicitly by\cite{aronson,NT}
\begin{align}
l^{\ast}  & =l+\frac{L}{r}\overline{m}+\frac{\overline{L}}{r}m+O(r^{-2}%
)\label{null rot}\\
m^{\ast}  & =m+O(r^{-1})\nonumber\\
n^{\ast}  & =n\nonumber
\end{align}
where $L=L(u,\zeta,\overline{\zeta})$ is a stereographic angle field given on
$\mathfrak{I}^{\mathfrak{+}}$ (still to be described) that determines the new
null geodesic field, $l^{\ast}.$

The transformation law for $\phi_{0\,i}^{0}$ (given only approximately for
small origin shifts\cite{explain}) is\cite{BIG}
\begin{equation}
\phi_{0\,i}^{\ast0}=(\phi_{0\,}^{0}-2L\phi_{1\,}^{0}+...)_{i}%
\label{transformation}%
\end{equation}

If we are given a Minkowski space world-line, $x^{a}=\xi^{a}(s),$ for the apex
of the new light cones, then $L=L(u,\zeta,\overline{\zeta})$ is given in the
parametric form
\begin{align}
L(u,\zeta,\overline{\zeta})  & =\xi^{a}(s)m_{a}(\zeta,\overline{\zeta
}),\label{L}\\
u  & =\xi^{a}(s)l_{a}(\zeta,\overline{\zeta}),\nonumber
\end{align}
with
\begin{align}
l_{a}(\zeta,\overline{\zeta})  & =\frac{\sqrt{2}}{2}(1,\frac{\zeta
+\overline{\zeta}}{1+\zeta\overline{\zeta}},-i\frac{\zeta-\overline{\zeta}%
}{1+\zeta\overline{\zeta}},\frac{-1+\zeta\overline{\zeta}}{1+\zeta
\overline{\zeta}}),\label{l}\\
m_{a}(\zeta,\overline{\zeta})  & =(0,Y_{1i}^{1}(\zeta,\overline{\zeta}%
))=\frac{\sqrt{2}}{2P}(0,1-\overline{\zeta}^{2},-i(1+\overline{\zeta}%
^{2}),\text{ }2\overline{\zeta}),\label{m}%
\end{align}

By the appropriate choice of $\xi^{a}(s),$ from Eq.(\ref{transformation}),
with the use of Eq.(\ref{L}), one can force the real part of $\phi_{0\,i}%
^{*0}$ to vanish, thereby making $x^{a}=\xi^{a}(s)$ the center of charge. If
however we generalized the choice of $L(u,\zeta,\overline{\zeta} )$ and
allowed it to be defined parametrically by%

\begin{align}
L(u,\zeta,\overline{\zeta})  & =\xi_{C}^{a}(\tau)m_{a}(\zeta,\overline{\zeta
}),\label{L*}\\
u  & =\xi_{C}^{a}(\tau)l_{a}(\zeta,\overline{\zeta}),\nonumber
\end{align}
where $z^{a}=\xi_{C}^{a}$ is a complex analytic world in complex Minkowski
space, then by setting $\phi_{0\,i}^{\ast0}=0,$ in Eq.(\ref{transformation})
the complex world-line is determined. This complex curve (which is purely
formal) defines the \textit{complex center of charge}. Using this "curve" as
the origin both the elecric \textit{and} magnetic dipoles vanish.

\textbf{Theorem\cite{footprints} - copy from home}

\section{The Complex Center of Mass}

For asymptotically flat Einstein-Maxwell space-times the situation is totally
analogous: the shear-free null geodesics originating from light-cones from
world-lines (real or complex) are replaced by (regular) asymptotically
shear-free null geodesic congruences generated by a complex
world-line\cite{footprints}, in the space of the complex Poincare translation
subgroup of the BMS group. The Maxwell asymptotic dipole transforms exactly as
in the flat space case, i.e., as in Eq.(\ref{transformation}) with however a
slight change in the parametric description of the function $L(u,\zeta
,\overline{\zeta}):$%

\begin{align}
L  & =\xi^{i}(\tau)Y_{1i}^{1}(\zeta,\overline{\zeta})-6\xi^{ij}(\tau
)Y_{2ij}^{1}(\zeta,\overline{\zeta}),\label{L***}\\
u  & =\frac{1}{\sqrt{2}}\xi^{0}(\tau)-\frac{1}{2}\xi^{i}(\tau)Y_{1i}^{0}%
(\zeta,\overline{\zeta})+\xi^{ij}(\tau)Y_{2ij}^{0}(\zeta,\overline{\zeta
})+...,
\end{align}
where the extra terms come from the existence of a non-vanishing Bondi shear,
given up to $l=2$ terms, by%

\begin{equation}
\sigma=24\xi^{ij}(u)Y_{2ij}^{2}+...\label{sigma}%
\end{equation}

Turning the gravitational behavior, the relevant (for us) tetrad components of
the Weyl tensor\cite{NT,BIG}%

\begin{align*}
\psi_{1}^{\,}  & =-C_{abcd}l^{a}m^{b}l^{c}m^{d}\\
\psi_{2}^{\,}  & =-C_{abcd}\overline{m}^{a}n^{b}l^{c}m^{d}%
\end{align*}
have the asymptotic form (the peeling theorem)
\begin{align*}
\psi_{1}^{\,}  & =\frac{\psi_{1}^{0\,}(u,\zeta,\overline{\zeta})}{r^{4}%
}+O(r^{-5})\\
\psi_{2}^{\,}  & =\frac{\psi_{2}^{0\,}(u,\zeta,\overline{\zeta})}{r^{3}%
}+O(r^{-4})
\end{align*}

The leading terms have the harmonic expansion:%

\begin{align}
\psi_{2}^{0\,}  & =\Upsilon+\psi_{2i}^{0}Y_{1i}^{0}+\psi_{2ij}^{0}Y_{2ij}%
^{0}+...\label{exp 4}\\
\psi_{1}^{0}  & =\psi_{1i}^{0}Y_{1i}^{1}+\psi_{1ij}^{0}Y_{2ij}^{1}%
+...\label{exp 6}%
\end{align}

The mass aspect, defined by
\[
\Psi=\psi_{2}^{0\,}+\eth^{2}\overline{\sigma}+\sigma(\overline{\sigma}%
)^{\cdot},
\]
is \textit{real} and has the expansion
\[
\Psi=\overline{\Psi}=\Psi^{0}+\Psi^{i}Y_{1i}^{0}+\Psi^{ij}Y_{2ij}^{0}+...
\]
The Bondi mass and linear momentum (four-momentum) is obtained from the
$l=(0,1)$ harmonic components of $\Psi$ by$:$%
\begin{align}
\Psi^{0}  & =-\frac{2\sqrt{2}G}{c^{2}}M\label{M}\\
\Psi^{i}  & =-\frac{6G}{c^{3}}P^{i}.\label{P^i}%
\end{align}
The \textit{complex gravitational dipole moment }(roughly mass dipole plus $i
$ times angular momentum) is identified as being proportional to the $l=1$
harmonic of $\psi_{1}^{0},$ i.e., as $\psi_{1i}^{0}.$ (Though many authors add
further terms to $\psi_{1i}^{0}$ for this identification, to our order of
approximation they all agree with our identification\cite{BIG}.)

The transformation (to second order) of $\psi_{1i}^{0}$ to an arbitrary
(complex) world-line, analogous to Eq.(\ref{transformation}), using
Eq.(\ref{L***}), is\cite{BIG}%

\begin{equation}
\psi_{1i}^{*0}=(\psi_{1}^{0}-3L\psi_{2}^{0}+...)_{i}.\label{transformation*}%
\end{equation}
Setting $\psi_{1i}^{*0}=0,$ thereby defining the complex center or mass,
$\xi^{i}(u),$ yields after a lengthy calculation,%

\begin{equation}
\psi_{1i}^{0}=-\frac{6\sqrt{2}G}{c^{2}}M[\xi^{i}(w)+i\frac{1}{2}\epsilon
_{kji}v^{k}\xi^{j}]+G^{i},\label{gr.dipole}%
\end{equation}
where $G^{i}$ is a known non-linear function of quadrupole terms.

At this point we make our only assumption on the physical system being
considered. We saw that we could determine a complex center of charge or a
complex gravitational center of mass by setting either $\varphi_{0i}^{*0}$ or
$\psi_{1i}^{*0}$ to zero. \textit{We now assume, for the rest of this work,
that the two complex world-lines coincide.} Aside from taking $Q\neq0, $ there
are no other conditions on the internal structure of our source (particle).

We now turn to the dynamics, which are contained in the asymptotic Bianchi
identities. They can be written:%

\begin{align}
(\psi_{1}^{0\,})^{\cdot}  & =-\text{$\eth$}\Psi+\text{$\eth$}\sigma
(\overline{\sigma})^{\cdot}+3\sigma\text{$\eth$(}\overline{\sigma})^{\cdot
}+\text{$\eth$}^{3}\overline{\sigma}\text{ }+2k\phi_{1}^{0}\overline{\phi}%
_{2}^{0}\label{1*}\\
\Psi^{\cdot}  & =\sigma^{\cdot}\overline{\sigma}^{\cdot}+k\phi_{2}%
^{0}\overline{\phi}_{2}^{0}\label{2*}\\
k  & =2Gc^{-4}\label{3*}%
\end{align}
\qquad

From these two equations, (\ref{1*}) and (\ref{2*}), we extract the equations
of motion with the radiation reaction term. Rather than going thru the details
(long with rather unattractive calculations) we will describe what we did in
words and then give the results. First we point out that we change the Bondi
$u$-variable to $w=\sqrt{2}uc^{-1},$ $w$ being the retarded time. Derivatives
with respect to $w$ are denoted by prime, ($^{\prime}$).

We first extract from Eq.(\ref{1*}) its $l=1$ part and then decompose it into
its real and imaginary parts. This yields two results: the imaginary part
determines the dynamics of the total angular momentum, i.e., the conservation
of angular momentum. Other than remarking that we identify $S^{i}=Mc\xi
_{I}^{i}$ as the intrinsic spin (with $\xi_{I}^{i}$ the imaginary part of
$\xi^{i})$, this is not our interest here and will not be discussed any
further. The real $l=1$ part can be solved for the linear momentum $P^{i}$
that was sitting in the $l=1$ part of $\Psi:$%

\begin{equation}
P^{k}=Mv_{R}^{k}-\frac{2Q^{2}}{3c^{3}}v_{R}^{k{\Large \,}\prime}%
+W^{k}.\label{P***}%
\end{equation}
This is a major result that come from our identification of the complex
centers of mass and charge. First of all we see \textit{kinematic} expressions
for the Bondi 3-momentum, the $mv$ term and then the \textit{radiation
reaction contribution to the momentum}. The $W$ contains further kinematic
terms involving spin and quadrupole interactions that are known but not
displayed here\cite{BIG}.

Extracting the $l=(0,1)$ harmonics from Eq.(\ref{2*}) yields the Bondi mass
and momentum loss equations:%

\begin{align}
M^{\,\prime}  & =-\frac{G}{5c^{7}}{\Large (}Q_{Mass}^{ij\,\prime\prime\prime
}Q_{Mass}^{ij\,\prime\prime\prime}+Q_{Spin}^{ij\,\prime\prime\prime}%
Q_{Spin}^{ij\,\prime\prime\prime}{\Large )}-\frac{2Q^{2}}{3c^{5}}%
(v_{R}^{i\,\prime}v_{R}^{i\,\prime}+v_{I}^{i\,\prime}v_{I}^{i\,\prime
})\label{M'}\\
& -\frac{1}{180c^{7}}{\Large (}D_{E}^{ij\,\prime\prime\prime}D_{E}%
^{ij\,\prime\prime\prime}+D_{M}^{ij\,\prime\prime\prime}D_{M}^{ij\,\prime
\prime\prime}{\Large )}\nonumber\\
P^{k\,\prime}  & =F^{k}\equiv\frac{2G}{15c^{6}}{\Large (}Q_{Spin}%
^{lj\,\prime\prime\prime}Q_{Mass}^{ij\,\prime\prime\prime}-Q_{Mass}%
^{lj\,\prime\prime\prime}Q_{Spin}^{ij\,\prime\prime\prime}{\Large )}%
\epsilon_{ilk}-\frac{Q^{2}}{3c^{4}}(v_{I}^{l\,\prime}v_{R}^{i\,\prime}%
-v_{R}^{l\,\prime}v_{I}^{i\,\prime})\epsilon_{ilk}\label{P'}\\
& +\frac{Q}{15c^{5}}{\Large (}v_{R}^{j\,\prime}D_{E}^{jk\,\prime\prime\prime
}+v_{I}^{j\,\prime}D_{M}^{jk\,\prime\prime\prime}{\Large )}+\frac{1}{540c^{6}%
}{\Large (}D_{E}^{lj\,\prime\prime\prime}D_{M}^{ij\,\prime\prime\prime}%
-D_{M}^{lj\,\prime\prime\prime}D_{E}^{ij\,\prime\prime\prime}{\Large )}%
\epsilon_{ilk}\nonumber
\end{align}
with the mass and spin quadrupoles related to the $\xi^{ij}$ by%

\begin{equation}
\xi^{ij}=(\xi_{R}^{ij}+i\xi_{I}^{ij})=\frac{G}{12\sqrt{2}c^{4}} (Q_{Mass}%
^{ij\prime\prime}+iQ_{Spin}^{ij\prime\prime}).\label{quadu}%
\end{equation}
The mass loss equation is thus exactly the usual quadrupole energy loss plus
the classical dipole and quadrupole electromagnetic energy loss.

It is however Eq.(\ref{P'}) that is of most interest to us. By substituting
the kinematic expression for the momentum, Eq.(\ref{P***}) into Eq.(\ref{P'})
we obtain our generalized Abraham-Lorentz equations of motion:
\begin{equation}
Mv_{R}^{k\prime}+v_{R}^{k}M^{\prime}-\frac{2Q^{2}}{3c^{3}}v_{R}^{k{\Large \,}%
\prime\prime}+R^{k}=F^{k}.\label{A.L.}%
\end{equation}

Note that though it is similar to the Abraham-Lorentz equations there are many
differences that are hidden in the known but complicated expressions for
$M^{\prime},$ $R^{k}\ $and $F^{k}.$ The $F^{k}$ is the Bondi recoil (or
rocket) force due to the momentum loss, while $R^{k}$ can be considered to be
a gravitational radiation reaction force depending on internal degrees of
freedom spin and quadrupole moments. $M^{\prime}$ has exactly the classical
dipole energy loss terms plus two additional terms. Though it is very hard to
directly see if the solutions to Eq.(\ref{A.L.}) are well behaved, in the
conclusion we will discuss this issue in more general terms.

\section{Conclusions}

We have considered the situation of a gravitating - electromagnetic source of
compact support viewed from future null infinity. The only restriction made on
the distributions is that the total charge is non-vanishing and that the
\textit{complex asymptotic electromagnetic dipole moment} be proportional the
\textit{complex gravitational dipole moment} so that the complex centers of
charge and mass coincide. Though it is not clear how severe this condition is,
it certainly is a serious restriction. It has been shown that for this
situation the gyromagnetic ration, the ratio of the spin-angular momentum to
the magnetic moment is that of Dirac's, namely $g=2$. We showed that in a
manner completely analogous to the flat space Maxwell case, one could
determine the transformation laws for the two dipole moments and thereby go to
the center of mass/charge, determining a unique complex world-line. Then,
using the Bianchi Identities, which play the role of dynamical equations, we
were able to give kinematic significance to the Bondi linear momentum, in the
sense that we had
\begin{equation}
\overrightarrow{P}=M\overrightarrow{v}-\frac{2Q^{2}}{3c^{3}}\overrightarrow
{v^{\prime}}+...\label{P.A.L.*}%
\end{equation}

From the Bondi momentum loss equation it immediately followed that we had a
generalized version of the Abraham-Lorentz equations of motion for an isolated
massive charged particle. It should be emphasized that the quadrupole
quantity, $\xi^{ij}(u),$ is arbitrary and in most cases it is taken as
non-vanishing in a finite interval so that the gravitational radiation exists
also in a finite interval. If however \textit{the motion, }from Eq.(\ref{A.L.}%
)\textit{, is unstable, }the particle acceleration will be unbounded and there
will be an infinite energy loss via the \textit{electromagnetic dipole
radiation}. This would be a physically unacceptable situation, indicating that
something is seriously wrong with the Einstein-Maxwell equations.

\quad The question then is does the general relativity (gravitational)
contributions to the equations of motion stabilize the equations. Though we do
not see any immediate prospects for a direct proof, we make a few comments.
Looking at Eq.(\ref{A.L.}), we see that the term $M^{\prime}v$ has the same
form as in the Abraham-Lorentz equation but now is more negative because of
the extra radiation terms and has the correct sign to try to stabilize the
motion. Whether or not it does stabilize is an open question. If not, perhaps
higher order terms that have been left out in our approximations might
succeed. And finally, it is known that the vacuum Einstein equations are
stable in the neighborhood of Minkowski space. If the same were true of the
Einstein-Maxwell equations with compact sources, that would constitute a proof
that our physical situation was indeed stable and the run-away behavior was
prevented by the inclusion of classical general relativity. Unfortunately,
this is a difficult question and, to our understanding, the answer is unknown.
It would be surprising if it turned out that the asymptotic Einstein-Maxwell
equations were unstable.

\section{Acknowledgements}

G.S.O. acknowledges the financial support from CONACYT and Sistema Nacional de
Investigadores (SNI-M\'{e}xico). C.K. thanks CONICET and SECYTUNC for support.


\begin{thebibliography}{99}
\bibitem {explain}The description of the finite transformation is considerably
more complicated.
\end{thebibliography}
\end{document}